\documentclass[twocolumn,10pt,secnumroman,showpacs,preprintnumbers,amsmath,amssymb]{revtex4}

\usepackage{graphicx}
\usepackage{dcolumn}
\usepackage{bm}
\usepackage{slashed}

\begin{document}

\title{\bf Angular invariant quantum mechanics in arbitrary dimension}

\author{ Sergio Giardino}
 \email{giardino@ime.unicamp.br}
\address{ Instituto de Matem\'{a}tica, Estat\'{i}stica e Computa\c{c}\~{a}o Cient\'{i}fica,
Universidade Estadual de Campinas\\
Rua S\'{e}rgio Buarque de Holanda 651, 13083-859, Campinas, SP,
Brazil}

\vspace{1cm}

\begin{abstract}
{\bf Resumo} Alguns dos problemas de mec\^{a}nica qu\^{a}ntica
unidimensional s\~{a}o generalizados em coordenadas esf\'{e}ricas e em
dimens\~{a}o arbitr\'{a}ria. S\~{a}o tratados os problemas do po\c{c}o de
potencial infinito, o oscilador harm\^{o}nico, a part\'{i}cula livre,
o potencial da fun\c{c}\~{a}o delta de Dirac, o po\c{c}o de potencial
finito e a barreira de potencial finito. As solu\c{c}\~{o}es da equa\c{c}\~{a}o
de Schr\"{o}dinger s\~{a}o escritas em termos das fun\c{c}\~{o}es de Bessel
e Whittaker e relacionadas a teorias f\'{i}sicas multi-dimensionais,
como a teoria de cordas.\linebreak

\vspace{5mm}

{\bf Abstract}
One dimensional quantum mechanics problems, namely the infinite potential well, the harmonic oscillator, the free particle, 
the Dirac delta potential, the finite well and the finite barrier are generalized for  finite arbitrary dimension in a radially
symmetric, or angular invariant, manner. This generalization enables the Schr\"{o}dinger equation 
solutions to be visualized for Bessel functions and Whittaker functions, and it also enables connections to multi-dimensional
physics theories, like string theory.

\vspace{5mm}

Palavras-chave: Mec\^{a}nica qu\^{a}ntica, fun\c{c}\~{o}es de Bessel, delta
de Dirac, f\'{i}sica em m\'{u}ltiplas dimens\~{o}es.

\end{abstract}
\noindent 


\maketitle

\section{Introduction}

An introductory quantum mechanics course deals with solutions to
one-dimensional problems, 
as can be seen in the commonly used textbooks on the subject. Three-dimensional
problems, like angular momentum, scattering and the hydrogen atom, are
not generalizations 
of one-dimensional problems, and they normally require either a
specific solution method to 
the Schr\"{o}dinger equation or an additional symmetry input. As a
multidimensional approach 
to one-dimensional problems does not necessarily lead to more
relevant multidimensional  
models, it is merely regarded as a curiosity. However, the importance of
multi-dimensional 
problems has increased in physics, since string theory has given rise to the
possibility 
that there could be more than three dimensions of space. For
example, the Schr\"{o}dinger 
equation was semi-classically solved in various dimensions in order to
quantize pulsating 
strings \cite{Minahan:2002rc,Giardino:2011jy}.

On the other hand, the more dimensions a problem has, the more
possibilities of motion, 
and the more symmetries it can have to restrict these
possibilities. This means that a generalization 
can be a choice, depending of the symmetries of the $n-$dimensional
solution of the problem. In this article we answer the question of what happens
when the Schr\"{o}dinger 
equation is solved using a formalism which generalizes
one-dimensional problems 
into angular invariant $n-$dimensional cases. Accordingly, the
Schr\"{o}dinger equation is 
transformed into the 
Bessel equation, whose solutions appear in physics problems which
either have cylindrical 
symmetry or spherical symmetry \cite{Spencer:1988ft,Fluegge:1998pq}.

The multidimensional approach provides a deeper understanding of
the physics 
of the one-dimensional problems, as well as the capabilities and
limitations of the mathematical 
apparatus. As some of these results may be scattered throughout
 literature, it is 
useful, for both students and researchers, to have them presented in
a single place. 

This article is organized as follows: in section two, the
infinite $n-$dimensional 
cylindrical quantum well is solved. In the third section, the quantum
harmonic oscillator is 
studied in various dimensions, and it is shown that this problem
can be described in 
terms of Bessel or Whittaker functions. In section four, the
free-particle is analyzed, requiring a 
particular Dirac delta function. Section five
deals with the Dirac delta 
potential. The finite well and the finite barrier are dealt with
in the sixth seventh sections, respectively. Finally, a brief
conclusion rounds off this article.

\section{the infinite square well}

The Schr\"{o}dinger equation can be written schematically as an
Eigenvalue equation
\begin{equation}
\big(\hat\Pi^2 + V \big)\Psi=\mathcal{E}\Psi \label{schr}
\end{equation}
where $\Psi$ is the wave-function, $\mathcal{E}$ is the energy, $V$ is
the potential and $\hat\Pi^2$ is the momentum
operator. Considering an $(n+1)-$dimensional space, the squared
momentum operator, expressed by means of spherical coordinates, depends on a
Laplacian operator that has a radial term and an angular term so that
\begin{equation}
\hat\Pi^2=-\frac{\hbar^2}{2m}\Big(\hat\nabla_r^2+\frac{1}{r^2}\hat\nabla_\theta^2\Big),
\end{equation}
where $\hat\nabla_{a=r,\,\theta}^2$ is the term of the Laplace operator
for the radial coordinate and for the angular coordinates. For a
radial-only dependent potential, the wave-function, expressed as  $\Psi(r,\,\theta) =
s(r)\,w(\theta)$, so that $w(\theta)$ has $n$ angular variables, splits
(\ref{schr}) into two equations, namely:
\begin{eqnarray}
&&\frac{1}{s}\hat\nabla_r^2s+\epsilon-v-\frac{M}{r^2}=0 \label{rad}\\
&&\frac{1}{w}\hat\nabla_r^2w=-M\label{ang}
\end{eqnarray}
\noindent where $\epsilon=2m\mathcal{E}/\hbar^2$, $v=2mV/\hbar^2$ and $M$ is a separation constant that is zero for $n = 0$.
Equation (\ref{ang}) can be solved in terms of $n-$dimensional spherical harmonics, which can be
found elsewhere \cite{Avery:1989pq,Bateman:1955ht}. In order to solve
the radial equation, it has been set
\begin{equation}
s=\frac{u(r)}{r^\nu}\label{u_r}
\end{equation}
and thus (\ref{rad}) becomes,
\begin{equation}
r^2u^{\prime\prime}+(n-2\nu)r\,u^\prime+\big[(\epsilon-v)r^2+\nu(\nu-n+1)-M\big]u=0.\label{split}
\end{equation}
The solution for (\ref{split}) depends on the particular potential $v$. For an infinite potential well,
the potential is
\begin{equation}
v=\left\{ 
\begin{array}{cc}
 0      &\qquad \mbox{if} \qquad r<R\\
 \infty &\qquad \mbox{if} \qquad r>R,\label{pot0}
\end{array}\right.
\end{equation}
and for a radial dependent-only wave-function, $M = 0$, the choice
$\nu=\frac{n-1}{2}$ leads to the Bessel equation,
\begin{equation}
r^2\,u^{\prime\prime}+ r\, u^{\prime}+\big(\epsilon\, r^2-\nu^2\big)\,u = 0.\label{bess}
\end{equation}
Thus, the wave-function is expressed in terms of Bessel functions
\begin{equation}
\Psi_n(r)=\frac{1}{r^\nu}\Big( a J_\nu\big(\sqrt{\epsilon}\,r\big)+b\,Y_\nu\big(\sqrt{\epsilon}\,r\big)\Big)\label{wfw}
\end{equation}
where a and b are integration constants. The choice of $\nu=\frac{n-2}{2}$ would lead to the spherical
Bessel equation, whose spherical Bessel functions, $j_\nu$ and $y_\nu$,
are related to the usual Bessel functions as,
\begin{equation}
j_\nu(x)=\sqrt{\frac{\pi}{2x}}J_{\nu+\frac{1}{2}}\qquad\mbox{and}\qquad y_\nu(x)=\sqrt{\frac{\pi}{2x}}Y_{\nu+\frac{1}{2}},
\end{equation}
which generates an identical wave-function, thus, (9) is indeed the most general solution to the
problem. As $Y_\nu(x)$ is divergent in $x = 0$ for $n > 0$, then $b =
0$ because otherwise the wave-function 
is not normalizable, as we will see in a moment. The potential is infinite at $r > R$, thus
$\Psi(R) = 0$. Defining $r = r_N^{(n)}$ as the $N-$th zero of $J_\frac{\nu-1}{
2}(r)$, the quantized energy is obtained from $\sqrt{\epsilon}\,R = r_N^{(n)}$,
and it is expressed as 
\begin{equation}
\mathcal{E}_N^{(n)}=\frac{\hbar^2}{2m}\left(\frac{r_N^{(n)}}{R}\right)^2.
\end{equation}
As $R$ is a free parameter and $n$ is fixed by the geometry, the more
excited the level of the energy, the more zeros the wave-function
in the interval $(0, R)$ has.

The wave-function is interpreted as a density of probability of finding
a quantum particle in the space, and the sum of all probabilities is defined to
be equal to one. The normalization is the
condition that warrants the probability of finding the particle to be one, namely
\begin{equation}\label{norma}
\intop_{\mathcal{V}}\,d\tau\,\Psi\,\Psi^*=1.
\end{equation}
The integral is calculated over the entire space $\mathcal{V}$ using the
complex conjugate $\Psi^*$ and the volume
element $d\tau$. For a three-dimensional space
parameterized in spherical coordinates, the well-known formula $d\tau=r^2\sin^2\theta\,dr\,d\phi$ applies.  Using
the general wave-function $\Psi(x)=\mathcal{N}\Phi(x)$, 
the normalization constant $\mathcal{N}$ ajusts the value of
(\ref{norma}) to one. Radial wave-functions, so that $\Psi=\Psi(r)$,
permit to integrate the angular terms of $d\tau$ and absorb them in the
normalization constant. Thus, an $(n+1)-$dimensional space in
spherical coordinates has the effective volume element $d\tau=r^n\,dr$. Finally, we calculate
the  normalizalized wave-functions using the integral
\begin{eqnarray}
&&\intop dr\,r\Big(J_\nu\big(\sqrt{\epsilon}\,r\big)\Big)^2=\\
&&=\frac{r^2}{2}\Big[\Big(J_\nu\big(\sqrt{\epsilon}\,r\big)\Big)^2 -J_{\nu+1}\big(\sqrt{\epsilon}\,r\big)\,J_{\nu-1}\big(\sqrt{\epsilon}\,r\big)
\Big],\;\nonumber
\end{eqnarray}
so that $\intop_0^\infty dr\,r^n|\Psi|^2=1$ implies the normalized wave-function
\begin{equation}
\Psi(r)=\frac{1}{R}\sqrt{\frac{2}{-J_{\frac{\nu+1}{2}}\big(\sqrt{\epsilon}\,r\big)\,J_{\frac{\nu-3}{2}}\big(\sqrt{\epsilon}\,r\big)}}\frac{J_{\frac{\nu+1}{2}}\big(\sqrt{\epsilon}\,r\big)}{r^{\frac{n-1}{2}}}
\end{equation}
The wave-function is rotationally invariant. In this sense, the
solution calculated above is not valid in the $n = 0$ case; the usual
one-dimensional $n = 0$ solution has anti-symmetrical states 
due to the negative values that the argument of the wave-function has
in this case, which are not included in a rotationally invariant
$(n+1)-$dimensional wave-function.
\section{the harmonic oscillator}
The one-dimensional harmonic oscillator is solved analytically in
terms of Hermite polynomials, as originally demonstrated by
Schr\"{o}dinger \cite{Schroedinger:1926oh}. This method uses the asymptotic 
behavior of the wave-function to simplify the problem and to obtain the
Hermite equation. However, using the variable $\rho= \mu\,r^2$ and the index $\nu= \frac{n+1}{2}$ in equation (\ref{split}) it is obtained
\begin{equation}
u^{\prime\prime}+\Big[-\frac{1}{4}+\frac{\epsilon}{4\,\mu\,\rho}-\frac{(n+1)(n-3)+4M}{16\rho^2}\Big]\,u=0\label{whitt}
\end{equation}
where $\mu=\frac{m\omega}{\hbar}$ and the prime means differentiation
relative to $\rho$. Equation (\ref{whitt}) with $M = 0$ is the Whittaker equation,
whose general solution is 
\begin{equation}
u(\rho)=a\,M_{\lambda,\,\eta}(\rho)+n\,W_{\lambda,\,\eta}(\rho)\label{wwf}
\end{equation}
where
$\lambda=\frac{\epsilon}{4\mu}=\frac{\mathcal{E}}{2\hbar\omega}$, $\eta=\pm\frac{n-1}{4}$ 
and $a$ and $b$ are integration constants. The Whittaker
functions $M_{\lambda,\,\eta}$ and $W_{\lambda,\,\eta}$ may be expressed as
\begin{eqnarray}
&&M_{\lambda,\,\eta}(\rho)=e^{-\frac{\rho}{2}}\rho^{\frac{1}{2}+\eta}\,M\Big(\frac{1}{2}+\eta-\lambda,\,1+2\eta,\,\rho\Big)\label{hyperM}\\
&&W_{\lambda,\,\eta}(\rho)=e^{-\frac{\rho}{2}}\rho^{\frac{1}{2}+\eta}\,U\Big(\frac{1}{2}+\eta-\lambda,\,1+2\eta,\,\rho\Big).\label{hyperU}
\end{eqnarray}
$M(p, q, z)$ and $U(p, q, z)$ are confluent hyper-geometric functions
known as Kummer functions. 
As always, wave-functions must be normalizable, and Kummer functions
diverge if the 
first index p is not a negative integer. Thus, normalizable
wave-functions are orthogonal 
polynomials of the order $N \in \mathbb{N}$, obtained through the following relations:
\begin{eqnarray}
&&
M\Big(-N,\,\frac{1}{2},\,z^2\Big)=\frac{(-1)^N\,N!}{\big(2N\big)!}\,H_{2N}(z)\label{relM1}\\
&&M\Big(-N,\,\frac{3}{2},\,z^2\Big)=\frac{(-1)^N\,N!}{\big(2N+1\big)!}\,\frac{H_{2N+1}(z)}{2z}\label{relM2}\\
&&U\Big(\frac{1-N}{2},\,\frac{3}{2},\,z^2\Big)=\frac{H_{N}(z)}{2^Nz}\label{relU1}\\
&&U\big(-N,\,\alpha+1,\,z\big)=(-1)^N\,N!\,L^{(\alpha)}_N(z)=\nonumber\\
&&=(-1)^N\,(\alpha+1)_N\,M\big(-N,\,\alpha+1,\,z\big)\label{relU2}
\end{eqnarray}
where $(\alpha + 1)_N$ is a Pochhammer symbol, $H_N(z)$ are the
Hermite polynomials, and $L^{(\alpha)}_N (z)$ 
are the generalized Laguerre polynomials. The one-dimensional solution
for the harmonic oscillator is obtained by setting $n = 0$. In this
situation, $b = 0$ in (\ref{wwf}), according to two reasons: 
when $N$ is even, (\ref{relU2}) is divergent in $z = 0$ and so the wave-function
is not normalizable, and when $N$ is odd, the wave-function is
normalizable but the energy integral \cite{Fluegge:1998pq}
\begin{equation}
\mathcal{E}=\intop_{-\infty}^\infty\,dr\,r^n\big|\hat\Pi\Psi\big|^2
\end{equation}
is divergent, thus solutions involving (\ref{relU1}) have to be discarded. The
remaining conditions (\ref{relM1}) and (\ref{relM2}) imposed on $\Psi$
in the $n = 0$ case, enables us to write the energy spectrum of the linear harmonic
oscillator
\begin{eqnarray}
&&\mathcal{E}_{2N}=\hbar\omega\Big(2N+\frac{1}{2}\Big)\\
&&\mathcal{E}_{2N+1}=\hbar\omega\Big(2N+\frac{3}{2}\Big)
\end{eqnarray}
Using the orthogonality condition for Hermite polynomials
\begin{equation}
\intop_{-\infty}^\infty\,dx\, e^{-x^2}\,H_L(x)\,H_K(x)=2^K\,K!\sqrt{\pi}\delta_{K,\,L}
\end{equation}
the normalized wave-function for the one-dimensional case is obtained with,
\begin{equation}
\Psi_K=\frac{1}{\sqrt{2^K\,K!}}\Big(\frac{\mu}{\pi}\Big)^\frac{1}{4}e^{-\frac{1}{2}\mu
  r^2}H_K\big(\sqrt{\mu}\,r\big)\label{psiho}
\end{equation}
where (\ref{psiho}) is valid for $K = 2N$ and $K = 2N + 1$. For the
$n-$dimensional case, (\ref{relU2}) indicates 
that the solution is given in terms of generalized Laguerre
polynomials, $L^{(\alpha)}_N $ and thus
$a = 0$ is established in (\ref{wwf}), without loss of
generality. When comparing (\ref{hyperU}) and (\ref{relU2}) we get
$\alpha= \pm\frac{n−1}{2}$ . As $n > 0$ and $\alpha>−1$, the plus sign
must be chosen. Also from (\ref{hyperU}) and (\ref{relU2}), we get the energy spectrum
\begin{equation}
\mathcal{E}_N=\hbar\omega\Big(2N+\frac{m+1}{2}\Big).
\end{equation}
Using the orthogonality relation
\begin{equation}
\intop_{0}^\infty\,dx\,x^\alpha
e^{-x}\,L_M^{(\alpha)}(x)\,L_N^{(\alpha)}(x)=\Gamma(1+\alpha)\binom{N+\alpha}{N} 
\delta_{M,\,N}
\end{equation}
the normalized wave-function is
\begin{equation}
\Psi_N(r)=\mu^{1/4}(-1)^N\sqrt{\frac{2\Gamma(N+1)}{\Gamma\Big(N+\frac{n+3}{2}\Big)}}\,e^{-\frac{1}{2}\mu
  r^2}L_N^{(\frac{n-1}{2})}(\mu r^2)
\end{equation}
The result shows that the energy depends explicitly on the angular dimension $n$ and that the
wave-function has a rotational symmetry, as expected from the angular independence
imposed by using $M = 0$ in the Schr\"{o}dinger equation.

\section{the free particle}
The equation that describes a free particle is similar to the equation
for the infinite well, as 
both have zero potential. The difference resides in the boundary
conditions. In the $n = 0$ 
case, the solution is expressed in terms of complex exponentials, and in
the arbitrary $n$ case, it is expressed in terms of Hankel functions,
which describe cylindrical travelling waves expressed in 
terms of Bessel functions, namely,
\begin{eqnarray}
&&H_\nu^{(1)}(z) = J_\nu(z) + i Y_\nu(z)\qquad \mbox{and}\qquad \\
&&H_\nu^{(2)}(z) = J_\nu(z) - i Y_\nu(z).
\end{eqnarray}
A general cylindrically symmetrical solution (\ref{u_r}) of (\ref{rad}) for the free
particle is,
\begin{equation}
\Psi_{\frac{n-1}{2}}(r)=\frac{1}{r^{\frac{n-1}{2}}}\Big(a\,H_{\frac{n-1}{2}}^{(1)}(\sqrt{\epsilon}
r)+b\,H_{\frac{n-1}{2}}^{(2)}(\sqrt{\epsilon}
r)\Big),\label{wffp}
\end{equation}
in which $H_\nu^{(1)}$  is a travelling mode towards $r = 0$, and $H_\nu^{(2)}$
 is a travelling mode towards $r\to\infty$ 
and $a$ and $b$ are integration constants. As in the $n = 0$ case, the
free-particle wave-function 
is not normalizable and it is understood as a wave packet which obeys,
\begin{equation}
\intop_0^\infty dr\,r^n\Psi_\eta^\dagger\Psi_\delta=\delta^{n+1}(\eta-\delta)\label{diracd}
\end{equation}
where $\delta^{n+1}(\eta-\delta)$ is an $(n+1)−$dimensional Dirac
delta function. By substituting two wave-functions 
like (\ref{wffp}) with energies $\eta$ and $\delta$ in (\ref{diracd}), and considering
\begin{eqnarray}
&&H_\nu^{(1)}(z)=\frac{1}{i\sin\nu\pi}\Big(J_{-\nu}(z)-e^{-i\nu\pi}J_\nu(z)\Big),\\
&& H_\nu^{(2)}(z)=\frac{1}{i\sin\nu\pi}\Big(-J_{-\nu}(z)+e^{i\nu\pi}J_\nu(z)\Big),\nonumber
\end{eqnarray}
and $J_{-\nu}=(-1)^\nu J_\nu$, we find that
\begin{equation}
\delta^{n+1}(\epsilon-\eta)=\sqrt{\epsilon}\intop_0^\infty dr\,r\,
J_\nu(\sqrt{\epsilon} r)\,J_\nu(\sqrt{\eta} r) 
\end{equation}
which is the definition of a Dirac delta function in terms of Bessel
functions, and thus 
the wave-function satisfies the mathematical requirements in order to
describe a cylindrically 
symmetric free-particle. However, one physical aspect is missing:
the behavior of the 
wave-function at $r = 0$. There is no external force or internal
interaction, thus at this point 
the travelling wave must change direction and maintain intensity. This means that the 
integration constants, which give the wave amplitude of the incoming
and outgoing waves, 
must have the same modulus in order to generate equal amplitudes for
the wave-function at $r = 0$. As $Y_\nu(0)$ is divergent and cannot
contribute to the solution, the wave-function is simply,
\begin{equation}
\Psi_{\frac{n-1}{2}}(r)=\frac{1}{r^{\frac{n-1}{2}}}\,J_{\frac{n-1}{2}}(\sqrt{\epsilon}r).
\end{equation}
Analogous to the $n = 0$ case, where wave-functions can be expressed
in terms of a Fourier 
transform which allows the free particle to be interpreted as a wave packet, the same can
be achieved here by expressing the free particle wave function as a Hankel
transform \cite{Boisvert:2010hb}
\begin{equation}
\Psi(r)=\intop_0^\infty d\sqrt{\epsilon}\,\phi(\sqrt{\epsilon})\,
J_\nu(\sqrt{\epsilon} r)\,(\sqrt{\epsilon} r)^{1/2}
\end{equation}
for an appropriate function $\phi(\sqrt{\epsilon})$. Thus, the analogy
between the $n = 0$ and the arbitrary $n$ 
is complete. Physically the general $n-$dimensional case is more
symmetrical because the 
wave-functions can only have one amplitude for both modes, something that does not
constrain the $n = 0$ one-dimensional case.

\section{the delta function potential}

In this case there is a Dirac delta function potential,
\begin{equation}
V=\pm g\,\delta(r-R)\label{delpot}
\end{equation}
in which $g > 0$ is the coupling constant of the potential. A negative
sign in (\ref{delpot}) means a 
potential well and a positive sign means a potential
barrier. Scattered states are possible 
for both signs of the potential, and a bound state occurs in the
potential well for negative 
energies, which is discussed in the following subsection.
\subsection{bound state}
The Schr\"{o}dinger equation with negative energy $\mathcal{E}
=-|\mathcal{E}|$ is expressed as 
\begin{equation}
\nabla^2\Psi+\gamma\delta(r-R)\Psi=\epsilon\Psi\label{schrdd}
\end{equation}
where $\gamma = \frac{2m}{\hbar^2} g$ and $\epsilon  = \frac{2m}{\hbar^2}|\mathcal{E}|$. The general solution to this problem is given in terms of
modified Bessel functions $I_\nu$ and $K_\nu$, integration constants $a$ and $b$, and $\nu =\frac{n-1}{2}$, so that
\begin{equation}
\Psi(r)=\frac{1}{r^\nu}\Big(a\,I_\nu(\sqrt{\epsilon}\,r)+b\,K_\nu(\sqrt{\epsilon}\,r)\Big).
\end{equation}
The modified Bessel function $I_\nu$ is divergent at $r\to\infty$ and
at $r\to0$, $K_\nu\to\infty$, thus the wave-function is,
\begin{equation}
\Psi(r)=\left\{
\begin{array}{cc}
\Psi_I=a\,\frac{I_\nu(\sqrt{\epsilon}\,r)}{r^\nu}&\qquad\mbox{if}\qquad
r<R\\
\Psi_{II}=b\,\frac{K_\nu(\sqrt{\epsilon}\,r)}{r^\nu}&\qquad\mbox{if}\qquad
r>R.
\end{array}
\right.
\end{equation}
At $r = R$, $\Psi_I=\Psi_{II}$ and one integration constant is eliminated,
\begin{equation}
a=\frac{K_\nu(\sqrt{\epsilon}\,R)}{I_\nu(\sqrt{\epsilon}\,R)}\,b.\label{intcte}
\end{equation}
On the other hand, the first derivative of the wave-function is not
continuous at $r = R$, as can be seen from integrating (\ref{schrdd})
in a $r = R$ neighborhood, which gives,
\begin{equation}
\Delta\big(r^n\Psi^\prime\big)=-\gamma\Psi(R).\label{deltaps}
\end{equation}
The $\epsilon\Psi$ term is eliminated from (\ref{schrdd}) by
integration, and this does not contribute to (\ref{deltaps}). 
This means that the energy sign of the energy is not important in order to determine whether the particle
is bound or free; all information regarding this is in the potential sign. At $r = R$,
\begin{eqnarray}
&&\Delta\big(r^n\Psi^\prime\big)=R^n\big(\Psi_{II}^\prime(R)-\Psi_I^\prime(R)\big)=\nonumber\\
&&=-R^{n-\nu}\sqrt{\epsilon}\big(a\,I_\nu(\sqrt{\epsilon}\,R)-b\,K_\nu(\sqrt{\epsilon}\,R)\big).\label{deltaprim}
\end{eqnarray}
Using (\ref{intcte}), (\ref{deltaps}), (\ref{deltaprim}), and the
Wronskian \cite{Boisvert:2010hb},
\begin{eqnarray}
&&K_\nu(x)\,I_{\nu+1}(x)+K_\nu(x)\,I_{\nu+1}(x)=\frac{1}{x},\nonumber\\
&&I_\nu(\sqrt{\epsilon}\,R)\,K_\nu(\sqrt{\epsilon}\,R)=\frac{1}{\gamma\,R}\qquad\mbox{is  obtained}.\label{relik}\nonumber
\end{eqnarray}
(\ref{relik}) is a transcendental equation and it enables us to
determine the energy numerically 
or graphically for each $n$. However, some particular cases can be calculated. For
\begin{eqnarray}
&& x\to\infty,\qquad K_\nu(x)\,I_{\nu}(x)\to\frac{1}{2x}\qquad\mbox{and
  thus}\qquad\nonumber\\ && \epsilon=\frac{\gamma^2}{4}\qquad\mbox{or}\qquad|\mathcal{E}|=\frac{mg^2}{2\hbar^2},
\end{eqnarray}
which is the unique bound state of this regime. On the other hand if
$x\ll 1$, then 
\begin{equation}
K_\nu(x)\,I_{\nu}(x)\to\frac{1}{2\nu}-\frac{x^2}{2\nu(\nu^2-1)}
\end{equation}
where $\nu>1$ and the energy for this regime is 
\begin{equation}
\epsilon=2\frac{\nu^2-1}{R}\Big(1-\frac{2\nu}{\gamma\,R}\Big),
\end{equation}
which is also a one$-$state solution only, as has been observed in the
well-known one-dimensional case.
\subsection{scattering state}
In this problem, the particle comes from infinity towards $r = 0$ and is scattered by a
Dirac delta well at $r = R$. In fact, the transmitted wave is totally
reflected at $r = 0$, thus in 
the $r < R$ region the waves have the same intensity in directions;
accordingly, the wave-function in 
\begin{eqnarray}
&&\Psi(r)=\\
&&=\left\{
\begin{array}{ll}
\Psi_I=a\,\frac{J_\nu(\sqrt{\epsilon}\,r)}{r^\nu}\qquad\mbox{if}\qquad
r<R&\\
\Psi_{II}=\frac{1}{r^\nu}\Big(b\,H^{(1)}_\nu(\sqrt{\epsilon}\,r)+H^{(1)}_\nu(\sqrt{\epsilon}\,r)\Big)\;\mbox{if}\; r>R
\end{array}
\right.\nonumber
\end{eqnarray}
so that $\nu=\frac{n-1}{2}$ and $H^{(2)}_\nu$ describes the incident wave. From the continuity of the wave
function and the integration of the Schr\"{o}dinger equation, we obtain
\begin{eqnarray}
&&a\,J_\nu-b\,H_\nu^{(1)}=H_\nu^{(2)}\\
&&a\,J_{\nu+1}+b\,\Big(\frac{\gamma}{\sqrt{\epsilon}}H^{(1)}_\nu+H^{(1)}_{\nu+1}\Big)=H^{(2)}_{\nu+1}-\frac{\gamma}{\sqrt{\epsilon}}H^{(2)}_\nu.\nonumber
\end{eqnarray}
All Bessel functions are evaluated at $\sqrt{\epsilon}\,R$. Using a
Wronskian for Hankel and Bessel 
functions, the above system can be solved for a and b, whose modulus
give us the reflection rate and the transmission rate, namely,
\begin{eqnarray}
&&R=|a|^2=1\qquad\mbox{and}\qquad \\
&&T=|b|^2=\frac{16}{\big(\pi\gamma R
  J_\nu Y_\nu\big)^2 +\big(\pi\gamma R J_\nu^2-2\big)^2 }.\nonumber
\end{eqnarray}
A reflection rate equal to one is understandable considering the fact
that at $r = 0$ the wave 
is totally reflected, and as the wave-function describe stationary
states, everything coming from infinite will be reflected. On the other hand, the
transmission rate is something altogether more subtle. It may be
greater than one, and if $R = 1$, it would be expected that $T = 0$. 
However, there is a reflection of the wave inside the region $ r \leq R$
at $r = R$, and thus it 
is understandable that, within this region, the intensity of the wave
will be greater than outside the region; the 
incoming wave is not immediately reflected to infinity and in fact
a stationary wave-function is
generated by totally reflection at the origin $r = 0$ and a partial transmission
at $r = R$. Thus, $T$ cannot be interpreted as a transmission of
the incoming wave, but as 
a relative intensity of the beams in the confined region and open
region. Inside each region, the incoming beam and the outgoing beam
have equal intensities. The intensity of the stationary wave drops to
zero if the position of the potential $R\to\infty$, but in the zeros
of the Bessel functions, it is four times greater than the incoming wave,
irrespective of how far the zero is from the origin of the coordinate system. 

The relative intensity of the wave-functions also enables energy quantization 
according to the value of $T$, which is an oscillating function. For each particular value of $T$,
there is an infinite spectrum of energy where the transmission has
this particular value. Thus, it can be said that the energy is
quantized for this system, because only particular values 
of the energy are permitted.

One last comment about this case must be made about the
wave-function. In the region $r > R$ 
there are incoming and outgoing waves represented by Hankel functions. The intensity of
these waves is equal, so $|b|^2 = 1$, although the coefficients are not
necessarily equal, thus, by ansatz, $b\neq 1$. In the $r < R$ region,
the situation is different, and the intensity and the Hankel functions
coefficients functions must be equal in both of the
directions because $Y_\nu\to\infty$ at $r\to 0$, something which does not
occur in the $r > R$ region,thus the greater generality of the
wave-function there. 

\subsection{Dirac delta function barrier}
The case of the Dirac delta function is totally analogous to the scattering of the Dirac
delta function well tackled above, the only difference being that the
sign of $\gamma$ sign in the potential term  is flipped from plus to minus. This
change, however, does not alter any of the results, which depend only
on $\gamma^2$, so the well and the barrier are physically indistinguishable.

A physical analogy of these models can be executed with a laser beam, which is produced
from an oscillating light-wave inside a partially reflecting
chamber. A light wave is produced 
inside the device and when the intensity of the wave inside it is
high enough, the coherent light escapes through one of the sides of
the chamber, which acts as a barrier, in a situation similar to the
Delta scattering. To be more realistic, the model would, of course, need
a source at $r = 0$. 

\section{the finite potential well}
In this case the potential is
\begin{equation}
v=\left\{
\begin{array}{l}
v_I=-v_0=-\frac{2m}{\hbar^2}\mathcal{V}_0\qquad\mbox{if}\qquad r<R\label{vfpw}
\\v_{II}=0\qquad\qquad\qquad\;\;\;\;\;\mbox{if}\qquad r>R
\end{array}
\right.
\end{equation}
where $\mathcal{V}_0>0$ and the potential describes a cylindrical well
whose center is located at $r = 0$. 
There are two possible solutions: a bounded-state solution with negative energy
and a scattering-state solution with positive energy.
\subsection{bound states}
This problem has negative energy $\mathcal{E} =-|\mathcal{E}|$ and
$\nu=\frac{n-1}{2}$ and $M = 0$ were chosen in (\ref{split}), thus obtaining,
\begin{equation}
r^2\,u^{\prime\prime}+r\,u^{\prime}+\big(\mathcal{Q}_a\,r^2-\nu^2\big)u=0,
\end{equation}
so that,
\begin{equation}
\mathcal{Q}_a=\left\{
\begin{array}{l}
\mathcal{Q}_I=v_0-|\epsilon|\qquad\mbox{if}\qquad r<R;\\
\mathcal{Q}_{II}=-|\epsilon|\;\;\;\,\qquad\mbox{if}\qquad r>R.
\end{array}
\right.
\end{equation}
The general solution to both regions is,
\begin{eqnarray}
&&u_I=a\,J_\nu\big(\sqrt{\mathcal{Q}_I}\,r\big)+b\,Y_\nu\big(\sqrt{\mathcal{Q}_I}\,r\big)\qquad\mbox{and}\\
&&u_{II}=c\,I_\nu\big(\sqrt{\epsilon}\,r\big)+d\,K_\nu\big(\sqrt{\epsilon}\,r\big).
\end{eqnarray}
The wave-function must be finite at $r = 0$ and at $r\to\infty$, thus
$b = c = 0$. The continuity of 
the wave-function and its first derivative at $r = R$ generates,
\begin{equation}
\frac{K_\nu\big(\sqrt{\epsilon}\,R\big)}{K_{\nu+1}\big(\sqrt{\epsilon}\,R\big)}\frac{J_{\nu+1}\big(\sqrt{\epsilon}\,R\big)}{J_\nu\big(\sqrt{\epsilon}\,R\big)}=\sqrt{\frac{|\mathcal{E}|}{\mathcal{V}_0-|\mathcal{E}|}}.
\end{equation}
This transcendental equation is solved numerically for each $n$, and
the intersection points of the graphs of both sides give us the
quantized energy. Quantized energy may be obtained for specific
cases. If $\mathcal{E}\approx\mathcal{V}_0$, then $J_\nu\to 0$, thus
the quantized energy comes from, $\sqrt{\mathcal{Q}_I}R = x_N^{(n)}$
\begin{equation}
\mathcal{E}_N=\mathcal{V}_0-\frac{2m}{\hbar^2}\left(\frac{x_N^{(\nu)}}{R}\right)^2,
\end{equation}
where $x^{(\nu)}_N$ is the $N−$th zero of $J_\nu$. In this situation
$R\gg x^{(\nu)}_N$ . If $n = 0$, Bessel functions
turn to trigonometric functions and the known regular spacing among these zeros for the
one dimensional well appears. Other quantizing possibilities come from
the $\mathcal{V}_0\gg\mathcal{E}$ regime. In this case the quantized
energy comes from the zero of $J_{\nu+1}$, namely
$\sqrt{\epsilon}\,R=x^{(\nu)}_N$ . This case can be understood as a
deep well or tiny energy. 
\subsection{scattering states}
In this case the state has positive energy and the potential is the
same used in the bounded states (\ref{vfpw}). As discussed in the case of the
free particle, the intensity of the wave-function is maintained at $r =
0$, and so the solution is simply, 
\begin{eqnarray}
&&\Psi(r)=\\
&&=\left\{
\begin{array}{l}
\Psi_I\,=\,b\,\frac{J_\nu\big(\sqrt{\mathcal{P}\,r}\big)}{r^\nu}\,\qquad\qquad\qquad\qquad\mbox{if}\qquad r<R;\\
\Psi_{II}=a\,\frac{H^{(1)}_\nu\big(\sqrt{\epsilon}\,r\big)}{r^\nu}+\frac{H^{(2)}_\nu\big(\sqrt{\epsilon}\,r\big)}{r^\nu}\qquad\mbox{if}\qquad
r>R,
\end{array}
\right.\nonumber
\end{eqnarray}
so that $\mathcal{P}=\epsilon+v_0$. From the continuity of the
wave-function and its first derivative, we obtain $|b|^2 = 1$ and,
\begin{eqnarray}
&&T=|a|^2=\\
&&=\frac{16/\big(\pi\,\epsilon\,R^2\big)}{\Big(\tilde
  J_\nu\,J_{\nu+1}-\mu\,J_\nu\,\tilde J_{\nu+1}\Big)^2+\Big(\tilde
  J_\nu\,Y_{\nu+1}-\mu\,Y_\nu\,\tilde J_{\nu+1}\Big)^2}\nonumber
\end{eqnarray}
where $\mu=\sqrt{\frac{\mathcal{V}_0}{\mathcal{E}}+1}$, $\tilde J_\nu = J_\nu(\sqrt{\mathcal{P}}\,R)$, $J_\nu =
J_\nu(\sqrt{\epsilon}\,R)$ and $Y_\nu =
Y_\nu(\sqrt{\epsilon}\,R)$. The result is compatible with the
situation for scattered states observed in the delta function model,
where the incoming wave is totally reflected at $r = R$ because
it is totally reflected at $r = 0$, and the intensity of the
wave-function is greater in the $r < R$ region because the outgoing
wave is partially transmitted at $r = R$. For each particular value of
$T$, which is an oscillating function, there is an infinite energy spectrum
that gives us this value, and then the energy is quantized 
according to this particular value. This result has also already been
obtained for the scattering in the Dirac potential.
\section{conclusion}
In this article several one-dimensional quantum mechanics problems have been generalized
in an angular invariant manner. The results confirm expectations such
as a contribution to the 
zero-point energy of the harmonic oscillator due to the angular
dimension, and several results are not 
so obvious: the equality of the wave-function intensities to
free particles, the existence of various quantized states in the Dirac
delta potential and the quantum scattering states for the finite
well. It is hoped that this set of results will be useful in
understanding quantum-mechanics problems that link angular invariance and
multi-dimensionality.

\paragraph* {\bf Acknowledgments} The author is grateful for the
facilities offered by the Mathematical Physics Department of the
University of S\~{a}o Paulo, and also for the financial support provided by Capes.
%
%
%
%
\bibliographystyle{unsrt}

\begin{thebibliography}{1}

\bibitem{Minahan:2002rc}
J.~A. Minahan.
\newblock {``Circular semiclassical string solutions on $AdS_5 x S_5$''}.
\newblock {\em Nucl.Phys.}, {\bf B648}:203--214, (2003) hep-th/0209047.

\bibitem{Giardino:2011jy}
S.~Giardino; V.~O. Rivelles.
\newblock {``Pulsating Strings in Lunin-Maldacena Backgrounds''}.
\newblock {\em JHEP}, {\bf 1107}:057, (2011) arXiv:1105.1353[hep-th].

\bibitem{Spencer:1988ft}
D.~E. Spencer;~P. Moon.
\newblock {``Field Theory Handbook''}.
\newblock Springer (1988).

\bibitem{Fluegge:1998pq}
S.~Fluegge.
\newblock {``Practical Quantum Mechanics''}.
\newblock Springer (1998).

\bibitem{Avery:1989pq}
J.~Avery.
\newblock {``Hyperspherical Harmonics''}.
\newblock Springer (1989).

\bibitem{Bateman:1955ht}
H.~Bateman.
\newblock {``Higher Transcendental Functions''}.
\newblock McGraw-Hill (1955).

\bibitem{Schroedinger:1926oh}
E.~Schroedinger.
\newblock {``Quantisierung als Eigenvertproblem''}.
\newblock {\em Ann.Phys.}, {\bf 79}:489, (1926).

\bibitem{Boisvert:2010hb}
R.~F. Boisvert; C. W. Clark; F. W. J. Olver; D.~W. Lozier.
\newblock {``NIST Handbook of Mathematical Functions''}.
\newblock Cambridge University Press (2010).

\end{thebibliography}

%
%
%
%
\end{document}